\documentclass[preprint,showpacs,showkeys,preprintnumbers,amsmath,amssymb,superscriptaddress]{revtex4}
\usepackage{graphicx}
\usepackage{dcolumn}
\usepackage{bm}

\begin{document}
\title{Gravitational energy of a noncommutative Vaidya black hole}

\author{S. Hamid Mehdipour}

\email{mehdipour@liau.ac.ir}

\affiliation{Department of Physics, Lahijan Branch, Islamic Azad
University, P. O. Box 1616, Lahijan, Iran}

\date{\today}

\begin{abstract}
In this paper we evaluate the components of the energy-momentum
pseudotensors of Landau and Lifshitz for the noncommutative Vaidya
spacetime. The effective gravitational mass experienced by a
neutral test particle present at any finite distance in the
gravitational field of the noncommutative Vaidya black hole is
derived. Using the effective mass parameter one finds that the
naked singularity is massless and this supports Seifert's
conjecture.

\end{abstract}


\maketitle

\section{\label{sec:1}Introduction}
There are several models of noncommutative geometry leading to an
effective grainy treatment of the spacetime manifold such that
coordinate operators might fail to commute \cite{land}. In 2005,
Nicolini {\it et al} \cite{nic1} in a physically inspired type of
noncommutativity based on the coordinate coherent state method,
via a minimal length induced by averaging noncommutative
coordinate fluctuations \cite{sma}, showed that the short distance
behavior of point-like structures can be cured. They derived the
metrics for some noncommutative black holes and found a new
thermodynamically stable final stage of the Hawking radiation in
which the curvature singularity of the black hole is removed
\cite{nic1,nic2}. Moreover, their method is consistent with
Lorentz invariance, unitarity and UV-finiteness of quantum field
theory. On the other hand, the mass scale associated with
noncommutativity, possibly and most reasonably is of the order of
the Planck scale. This is supported by the fact that the
fundamental Planck scale in models with large extra dimensions
becomes as small as a TeV in order to solve the hierarchy problem.
In addition, most of the phenomenological studies of the
noncommutative models have assumed that the mass scale associated
with noncommutativity cannot lie far above the TeV scale
\cite{hin}. Therefore the hope is that if the noncommutativity
scale is in the TeV range one might be able to detect some
evidences for it at future particle colliders \cite{riz}.

Recently, we investigated some aspects of the black hole
thermodynamics by using the noncommutative geometry inspired
formalism (the coordinate coherent state method) \cite{meh1}. In
this formalism the point-like particle, instead of being totally
localized at a point via a Dirac-delta function distribution, is
described as a smeared-like particle via a Gaussian distribution
of minimal width $\sqrt{\sigma}$, where $\sigma$ is the smallest
fundamental unit of an observable area in the noncommutative
coordinates, beyond which coordinate resolution is vague. As a
matter of fact, because of the appearance of high energies at
small scales of a noncommutative manifold, the influences of
manifold quantum fluctuations turn out to be apparent and forbid
any measurements to observe a particle location with a precision
more than an intrinsic length scale. To clarify more features on
these topics, see \cite{nic3} and the references included.

In the context of black hole physics, a non-static and spherically
symmetric spacetime is dependent upon an arbitrary dynamical mass
function. In this manner, an important issue is raised in
connection to how black hole mass decreases as a back-reaction of
the Hawking radiation, and thus it may be appropriately presented
by the Vaidya solution \cite{vai,lind}. The Vaidya black hole is
considered as a kind of a more applicable one owing to its
time-dependent decreasing mass. In this paper, we use the
noncommutative Vaidya (NCV) metric derived in Ref.~\cite{meh2} to
find its gravitational energy. There are several prescriptions for
computing the energy for a non-static general relativistic system
\cite{Ein,Tol,Pap,Lan,Ber,Gol,Mol,Wei}. For instance, one can use
Landau and Lifshitz's (LL) definition of energy \cite{Lan}, which
is easier to work out. Since a neutral test particle situated at
any finite radial distance in the gravitational field of the NCV
black hole experiences an effective gravitational mass, so this is
the subject of interest to us to evaluate the effective mass in
this field.

The paper is organized in the following. In Sec.~\ref{sec:2}, we
briefly discuss concerning the energy-momentum localization in the
literature. The energy-momentum distributions associated with the
NCV black hole are evaluated in Sec.~\ref{sec:3} and the LL
definition of energy is employed for these computations. The
summary is presented in Sec.~\ref{sec:4}. In this work, Greek
(Latin) indices run from 0 to 3 (1 to 3) and we use the natural
units with the following definitions: $\hbar = c = G = 1$.

\section{\label{sec:2}Energy-momentum localization}
The energy-momentum localization is one of the most significant
questions which survives uncertain in the theory of Einstein
general relativity (GR). There are plenty of attempts to attain a
specified approval for the explanation of energy-momentum as an
important conserved quantity in GR. Unfortunately, there is still
no generally confirmed explanation of energy and momentum
distributions in the literature \cite{Xul}. The energy-momentum
conservation in the context of GR is given by
\begin{equation}
\label{mat:1} \nabla _{\alpha}T^ {\alpha}_ {\beta} = 0,
\end{equation}
where $T^{\alpha}_{\beta}$ shows the symmetric energy-momentum
tensor containing the matter and all non-gravitational fields. An
expression for energy-momentum complexes due to including the
contribution from the gravitational field energy has been found by
Einstein \cite{Ein}. This expression is not a tensor and is
defined as the gravitational field pseudotensor
$t_{\beta}^{\alpha}$. The energy-momentum complex obeys the local
conservation laws as follows:
\begin{equation}
\label{mat:2}{\cal{T}}^{\alpha}_{\beta \,\, , \alpha}\equiv
\frac{\partial}{\partial x^{\alpha}} \left(\sqrt{-g}
(T_{\beta}^{\alpha} +t_{\beta}^{\alpha})\right) =0,
\end{equation}
where $g$ is the determinant of the metric tensor
$g_{\alpha\beta}$. The energy-momentum complex
${\cal{T}}^{\alpha}_{\beta}$ is substituted for the
energy-momentum tensor $T^{\alpha}_{\beta}$ which is a mixture of
$T^{\alpha}_{\beta}$ and $t_{\beta}^{\alpha}$ but in the usual
form of conservation laws. Then, we can write
\begin{equation}
\label{mat:3}{\cal{T}}^{\alpha}_{\beta}=\theta^{\alpha\gamma}_{\beta
\,\,\,\, , \gamma},
\end{equation}
where $\theta^{\alpha\gamma}_{\beta}$ are marked like
superpotentials which are not uniquely defined. With a suitable
choice of coordinates, one can vanish the pseudotensor
$t_{\beta}^{\alpha}$ at a special point, e.g. Schrodinger showed
that the pseudotensor can be disappeared outside the schwarzschild
radius. Numerous works have been accomplished by using a more
appropriate quantity to exhibit the energy and momentum
distributions of various gravitational backgrounds. The different
descriptions for the energy-momentum complexes
\cite{Ein,Tol,Pap,Lan,Ber,Gol,Mol,Wei} can only provide important
results if the calculations are performed in Cartesian
coordinates. Penrose \cite{Pen} introduced a method of quasi-local
energy to obtain the energy-momentum of a curved spacetime via
applying any coordinate system. Many research workers \cite{Vir}
have regarded a class of various suggestions of the quasi-local
energy to investigate various models of the universe. Nevertheless
the formalism of energy-momentum complexes could not produce some
unique explanation of energy in the framework of GR due to the
fact that each of these quasi-local expressions have their own
topics. In 1996, it was shown that the deferent energy-momentum
complexes coincide for any Kerr-Schild class metric and they give
the same energy distribution \cite{Vir2} in which their outcomes
are similar to the results of Penrose \cite{Pen} and also Tod
\cite{tod} utilizing the notion of quasi-local mass. Many
universal outcomes of the most general non-static spherically
symmetric spacetime by utilizing the Kerr-Schild class metric was
found afterwards by Virbhadra \cite{Vir3}. In 1999, Chang {\it et
al} \cite{chan} demonstrated that every energy-momentum complex is
connected to a Hamiltonian boundary term. Hence, the
energy-momentum complexes are quasi-local and suitable. It would
be worthwhile to denote that, in the next section we carry out the
calculations in quasi-Cartesian coordinates due to the fact that
the energy associated with the NCV metric using the LL
energy-momentum complex provides a meaningful result if
quasi-Cartesian coordinates are chosen \textbf{(for instance, see
\cite{Vir4})}.

\section{\label{sec:3}Energy-momentum distributions in NCV spacetime}
Since the dynamics for the black hole mass is a significant
problem, we therefore focus on a NCV spacetime which exhibits a
particularly rich dynamical structure. According to the
noncommutative geometry inspired method, one can obtain the NCV
metric in the presence of a smeared mass source by solving
Einstein equations \cite{meh2}. Let us first consider the diagonal
form of the NCV metric with respect to $\{x^\alpha\}=\{t, r,
\theta, \phi\}$ coordinates as follows:
\begin{equation}
\label{mat:4}ds^2=\left(1-\frac{2{\cal{M}}_\sigma}{r}\right)dt^2-\left(1-\frac{2{\cal{M}}_\sigma}{r}\right)^{-1}dr^2-r^2d\Omega^2,
\end{equation}
where $d\Omega^2=d\theta^2+\sin^2\theta d\phi^2$ is the line
element on the 2-dimensional unit sphere and ${\cal{M}}_\sigma$ is
the gaussian-smeared mass distribution given by
\begin{equation}
\label{mat:5}{\cal{M}}_\sigma=M_I\left[
{\cal{E}}\left(\frac{r-t}{2\sqrt{\sigma}}\right)\left(1+\frac{t^2}{2\sigma}\right)
-\frac{r}{\sqrt{\pi\sigma}}e^{-\frac{(r-t)^2}{4\sigma}}\left(1+\frac{t}{r}\right)\right],
\end{equation}
where $M_I$ is the initial black hole mass and ${\cal{E}}(n)$ is
the Gauss error function defined as ${\cal{E}}(n)\equiv
2/\sqrt{\pi}\int_{0}^{n}e^{-p^2}dp$. In the static case, $t=0$,
one regains the same result as Ref.~\cite{nic1}, i.e. the
noncommutative Schwarzschild metric and in the limit
$\sigma\rightarrow 0$ with $t=0$, one recovers the ordinary
Schwarzschild solution. We require to reexpress the NCV metric in
quasi-Cartesian coordinates. Transforming (\ref{mat:4}) to
Cartesian terms according to $x= r \sin\theta\cos\phi$, $y= r
\sin\theta \sin\phi$, and $z=r\cos\theta$, one gets the metric
\begin{equation}
\label{mat:6}ds^2=\left(1-\frac{2{\cal{M}}_\sigma}{r}\right)dt^2-dx^2-dy^2-dz^2-\frac{2{\cal{M}}_\sigma}{r^3-2{\cal{M}}_\sigma
r^2}\left(xdx+ydy+zdz\right)^2,
\end{equation}
where $r^2=x^2+y^2+z^2$. Now we are ready to express Einstein's
equations as defined by LL \cite{Lan}
\begin{equation}
\label{mat:7} L^{\alpha\beta} = \frac{1}{16\pi}{\cal
S}^{\alpha\beta\gamma\delta} _ {\,~~\quad ,\gamma\delta},
\end{equation}
where $L^{\alpha\beta}$ is the energy-momentum complex and
incorporates both the stress-energy tensor of matter and an
expression quadratic in first derivatives of the metric which is
symmetric with respect to its indices. The right-hand side, namely
LL superpotentials ${\cal S}^{\alpha\beta\gamma\delta}$, includes
an expression in which it has symmetries similar to the curvature
tensor and can be written as
\begin{equation}
\label{mat:8}{\cal S}^{\alpha\beta\gamma\delta} =
-g\left(g^{\alpha\beta}g^{\gamma\delta} -
g^{\alpha\gamma}g^{\beta\delta}\right).
\end{equation}
The LL energy-momentum complex satisfies the local conservation
laws
\begin{equation}
\label{mat:9} L^{\alpha\beta}_{~~\, , \beta} = 0.
\end{equation}
The $L^{00}$ is the energy density and $L^{0i}$ are the momentum
density components. The energy and momentum in the LL prescription
for a four-dimensional background are given by
\begin{equation}
\label{mat:10} P^\alpha=\int\int\int L^{0\alpha}dx^1dx^2dx^3.
\end{equation}
These integrals are extended over all space for $t = const$, and
have to be limited to the utilize of quasi-Cartesian coordinates
to have the meaning of energy and momentum. In order to evaluate
the energy and momentum distributions, we firstly have to compute
the LL superpotentials. There are seventy two nonzero
superpotentials in the LL prescription for the NCV spacetime but
the required ones are the following
\begin{equation}\label{mat:11}
\begin{array}{l}
{\cal S}^{1010} =-{\cal S}^{0011}=\frac{r^3-2{\cal{M}}_\sigma
x^2}{r^3-2{\cal{M}}_\sigma r^2}, \\
{\cal S}^{2010} ={\cal S}^{1020}=-{\cal S}^{0021} =-{\cal
S}^{0012}=-\frac{xy{\cal{M}}_\sigma}{r^3-2{\cal{M}}_\sigma
r^2}, \\
{\cal S}^{3010} ={\cal S}^{1030}=-{\cal S}^{0031} =-{\cal
S}^{0013}=-\frac{xz{\cal{M}}_\sigma}{r^3-2{\cal{M}}_\sigma
r^2}, \\
{\cal S}^{2020} =-{\cal S}^{0022}=\frac{r^3-2{\cal{M}}_\sigma
y^2}{r^3-2{\cal{M}}_\sigma r^2},\\
{\cal S}^{3020} ={\cal S}^{2030}=-{\cal S}^{0032} =-{\cal
S}^{0023}=-\frac{yz{\cal{M}}_\sigma}{r^3-2{\cal{M}}_\sigma
r^2},\\
{\cal S}^{3030} =-{\cal S}^{0033}=\frac{r^3-2{\cal{M}}_\sigma
z^2}{r^3-2{\cal{M}}_\sigma r^2}.\\
\end{array}
\end{equation}
Substituting LL superpotentials into Eq.~(\ref{mat:7}), the energy
and momentum densities take the form
\begin{equation}\label{mat:12}
\begin{array}{l}
L^{00} =
\frac{1}{4\pi(r-2{\cal{M}}_\sigma)^2}\left[\frac{{\cal{M}}_\sigma}{r}\left(\frac{9(r-2{\cal{M}}_\sigma)^2}{r^2}-1\right)+{\cal{M'}}_\sigma\right], \\
L^{01} = -\frac{2x{\cal{\dot{M}}}_\sigma}{8\pi
r(r-2{\cal{M}}_\sigma)^2}, \\
L^{02} = -\frac{2y{\cal{\dot{M}}}_\sigma}{8\pi
r(r-2{\cal{M}}_\sigma)^2}, \\
L^{03} = -\frac{2z{\cal{\dot{M}}}_\sigma}{8\pi
r(r-2{\cal{M}}_\sigma)^2}.\\
\end{array}
\end{equation}
The prime abbreviates $\partial/\partial r$, and the overdot
abbreviates $\partial/\partial t$. The quantities
${\cal{M'}}_\sigma$ and ${\cal{\dot{M}}}_\sigma$ are given by
\begin{equation}
\label{mat:13}{\cal{M'}}_\sigma=\frac{M_Ir^2}{2\sqrt{\pi\sigma^3}}\,e^{-\frac{(r-t)^2}{4\sigma}},
\end{equation}
\begin{equation}
\label{mat:14}{\cal{\dot{M}}}_\sigma=\frac{M_I}{\sqrt{\pi\sigma}}\left[
\sqrt{\frac{\pi}{\sigma}}\,t\,{\cal{E}}\left(\frac{r-t}{2\sqrt{\sigma}}\right)
-\left(\frac{r^2}{2\sigma}+2\right)e^{-\frac{(r-t)^2}{4\sigma}}\right].
\end{equation}
We are interested in computing the energy-momentum distributions
associated with the NCV black hole background, which are contained
in a sphere of radius $r_0$. Therefore, if we replace
Eqs.~(\ref{mat:12}) into Eq.~(\ref{mat:10}), applying the Gauss
theorem and evaluating the integrals over the surface of
two-sphere of radius $r_0$, we get the energy and momentum
components in the following form
\begin{equation}
\label{mat:15} E(r_0)=M_I-\frac{{\cal{M}}_\sigma(r_0)
r_0}{r_0-2{\cal{M}}_\sigma(r_0)},
\end{equation}
\begin{equation}
\label{mat:16}P^1=P^2=P^3=0.
\end{equation}
Note that the asymptotic value of the total gravitational mass of
a NCV black hole is the initial black hole mass $M_I$. The energy
distribution derived here is indeed the energy of the
gravitational field that a neutral particle experiences at a
finite distance $r_0$. Hence, the energy given by
Eq.~(\ref{mat:15}) can also be identified as the effective
gravitational mass $M_{eff}$ of the spacetime under consideration.
Many authors have devoted a worthwhile attention to the problem of
finding the effective mass for various spacetimes \cite{vag}.
Studying on this issue was first considered by Cohen and Gautreau
\cite{coh}. They were the first who introduced the idea of
effective mass by utilizing Whittaker's theorem and derived the
effective mass for the Reissner-Nordstr\"{o}m and Kerr-Newman
spacetimes.

In addition, the result in Eq.~(\ref{mat:15}) may be worthwhile
for testing Seifert's conjecture in the context of the naked
singularity arising from the spherical collapse described by the
metric (\ref{mat:4}). In 1979, Seifert \cite{sei} conjectured that
any singularity which appears is invisible if a finite nonzero
mass collapses into a point, or is naked if either one has
singularities along lines (or surfaces) or the central
singularities are carefully arranged that they include only zero
mass. In Ref.~\cite{meh2}, we analyzed the metric (\ref{mat:4}) in
three possible causal structures and found a zero remnant mass at
long times, i.e. an instable black hole remnant. So, a naked
singularity at $r=0$ in this non-static case is appeared which is
clear from Eq.~(\ref{mat:15}) that at the origin, the effective
mass of the NCV black hole is equal to zero. This supports the
Seifert conjecture. The naked singularity forming in the Vaidya
null dust collapse is also massless \cite{Vir3}, providing a
support to Seifert's hypothesis which approves our outcome.

Here, it is important to stress that the energy density given by
the first of the expressions in Eq.~(\ref{mat:12}) becomes
infinite when we set $r$ equal to the horizon radius $r_H$, so
that $r_H=2{\cal{M}}_\sigma(r_H)$. As a result, the energy
distribution becomes infinite for the case of the black hole
horizon. However, the condition of the infiniteness for the energy
density of a closed system which is confined to a two-sphere of
radius $r_0$ is not a sensible condition, thereby, this value for
the radial coordinate should be eliminated. On the other hand,
when one considers the time-varying mass in the NCV black hole,
one could possibly deal with this singularity at long times.
According to Ref.~\cite{meh2}, as time moves forward the minimal
nonzero mass decreases but the minimal nonzero horizon radius
increases which means that in the long-time limit, the horizon
radius tends to the infinity.

\section{\label{sec:4} Summary}
In summary, we have evaluated the energy and momentum
distributions in the NCV black hole background using the LL
prescription. The energy distribution of the NCV black hole that
is contained in a sphere of radius $r_0$ shows that a neutral test
particle situated at a finite distance $r_0$ experiences the
gravitational field of the effective gravitational mass in the
spacetime under study, while the momentum distributions $P^i$ are
equal to zero which are acceptable results. Using the expression
for energy we have found that the naked singularity forming in the
NCV black hole is massless and this confirms the Seifert
conjecture.

\end{document}